\documentclass[9pt,twocolumn,twoside]{opticajnl}
\journal{opticajournal} 

\setboolean{shortarticle}{true}

\usepackage[font=small,justification=justified]{caption}

\usepackage{lineno}

\title{Delay Mismatches Shape the Spectra of Three Synchronized Semiconductor Lasers}

\author[1,*] {Sergi Torres-Serra}
\author[2]{Mohammadreza Mokhtari}
\author[1]{Cristina Masoller}
\author[3,4]{Jordi Tiana-Alsina}
\author[2]{Apostolos Argyris}
\author[2]{Miguel C. Soriano}

\affil[1]{Department of Physics, Universitat Politècnica de Catalunya, Rambla Sant Nebridi 22, 08222 Terrassa, Barcelona, Spain}
\affil[2]{Institute of Cross-Disciplinary Physics and Complex Systems, IFISC (CSIC-UIB), Campus UIB, 07122 Palma de Mallorca}
\affil[3]{Department of Applied Physics, Faculty of Physics, Universitat de Barcelona, Martí i Franquès 1, 08028 Barcelona, Spain}
\affil[4]{Institut de Nanociència i Nanotecnologia (IN2UB), Universitat de Barcelona, 08028 Barcelona, Spain}

\affil[*]{sergi.torres.serra@upc.edu}

\begin{abstract}
\fontsize{10}{10.5}\selectfont
We study a network of three synchronized chaotic lasers and the radio frequency spectra of the intensity fluctuations. The position and morphology of the peaks are significantly different depending on the set of lasers being synchronized. The lasers and coupling paths are similar but not identical in all parameters, including their feedback and coupling delay times. The inverse of the delay mismatch between each pair of lasers introduces characteristic frequencies that shape the spectral response of the synchronized system. By studying two different three-laser configurations, we find that peaks appear whenever multiple harmonic contributions overlap in frequency, resulting in constructive enhancement of the spectral power. Numerical simulations performed with the same delay configurations show a clear agreement with the experimental measurements, reproducing both the peak positions and the overall spectral structure. These results indicate that the frequency content of the synchronized dynamics is largely determined by the delay-mismatch network and its harmonics.

\end{abstract}

\setboolean{displaycopyright}{false} 

\begin{document}

\maketitle

\fontsize{9}{10.8}\selectfont
Synchronization in chaotic systems has been extensively studied over the past several decades due to its fundamental interest and its potential applications in secure communications \cite{OhtsuboBook2017}, sensing \cite{Argyris2016}, and complex networks \cite{Pecora1990,Boccaletti2002}. Semiconductor lasers constitute a particularly attractive platform for investigating chaotic synchronization because of their inherently fast dynamics, which occur on sub-nanosecond timescales. As a consequence, the finite propagation time of flight between coupled lasers cannot be neglected for coupling distances beyond a few centimeters and must be explicitly taken into account through delayed coupling terms \cite{OhtsuboBook2017,Uchida2012,Heil2001,Wunsche2005,yanchuk2017spatio,tavakoli2020multi}. Under appropriate conditions, such as optical feedback or optical injection, semiconductor lasers can exhibit highly complex chaotic dynamics \cite{Fischer1996}. A wide variety of coupling architectures, including unidirectional, bidirectional, mutual, and network configurations, have been proposed to achieve synchronization between chaotic lasers \cite{Soriano2013}. Exploiting this synchronization capability, networks of mutually coupled semiconductor lasers provide a versatile platform for investigating collective phenomena, spatiotemporal dynamics, and a broad range of synchronization regimes arising from the interplay between nonlinear dynamics, coupling topology, and time delays \cite{Vicente2008, Arenas2008, Soriano2013, Flunkert2012}. 

\begin{figure}[ht]
\centering
\includegraphics[width=1.02\linewidth]{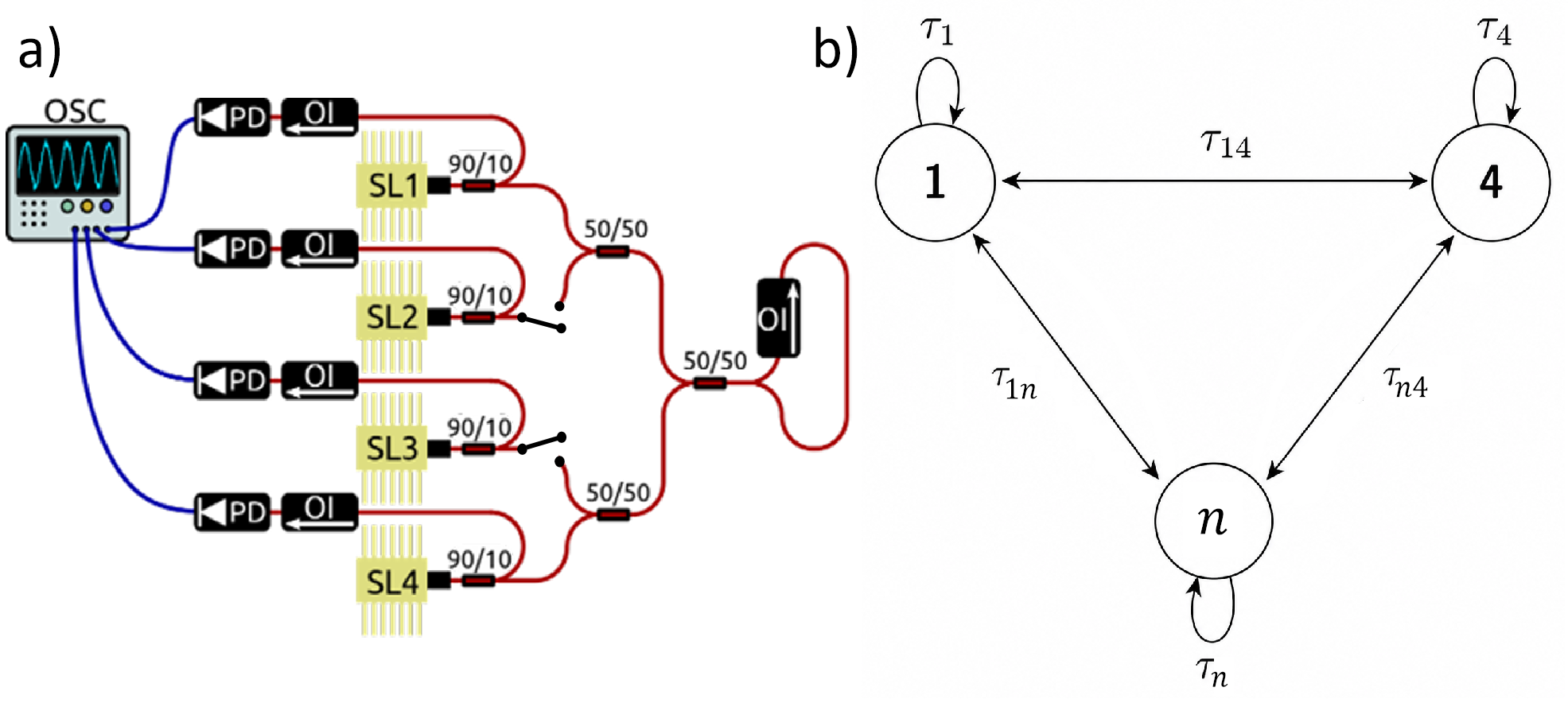}
\vspace{-0.7cm}
\caption{a) Experimental setup of the synchronized semiconductor laser network coupled through polarization-maintaining optical fibers. The setup includes two ON/OFF switches that allow selecting either laser 2 or laser 3, thereby enabling the 124 and 134 three-laser configurations. OSC: oscilloscope; PD: photodetector; OI: optical isolator; SL$n$: semiconductor laser; 90/10 and 50/50: optical couplers. (b) Topology of the corresponding three-node laser network, including mutual optical coupling and self-feedback. $\tau_i$ is the feedback delay time, and the coupling delay time is $\tau_{ij}=(\tau_i+\tau_j)/2$. The index $n$ denotes either laser 2 or laser 3.}
\label{fig:setup}
\vspace{-.2cm}
\end{figure}

While delay mismatch is known to modify the dynamics of coupled lasers \cite{Hicke2011}, its role remains elusive. Here we analyze the delay mismatches in a three-laser network, focusing on the RF spectral peaks. Fig.~\ref{fig:setup}a) presents the experimental setup, which is built such that we can test several three-laser networks. The four lasers we employed are ultra–low-threshold quantum-well distributed feedback (QW-DFB), subjected to optical delays, $\tau_i$, of 82.25~ns, 82.60~ns, 82.71~ns, and 82.88~ns, and are labeled as lasers 1 to 4 according to increasing delay. The feedback delay time for each laser was obtained from the first peak of the intensity autocorrelation function, with the laser subject to its own optical feedback while the other lasers were turned off. Here, all lasers are operated at a bias current of 1.2$I_{th}$. Either laser 2 or laser 3 is switched off to select the 134 or 124 three-laser configuration, respectively, both represented by the network topology in Fig.~\ref{fig:setup}b). The lasers are interconnected through polarization-maintaining optical fibers. Optical beam splitters are used both to distribute the optical signals among the lasers and to redirect a fraction of the emitted light to 20~GHz bandwidth photodetectors (PDs), whose electrical outputs are subsequently monitored using a high-speed oscilloscope (OSC). The RFS are obtained by the FFT of the time series, with a frequency range of 0~-~128~GHz and a resolution of 1~MHz.

\begin{figure}[ht]
\centering
\includegraphics[width=1.015\linewidth]{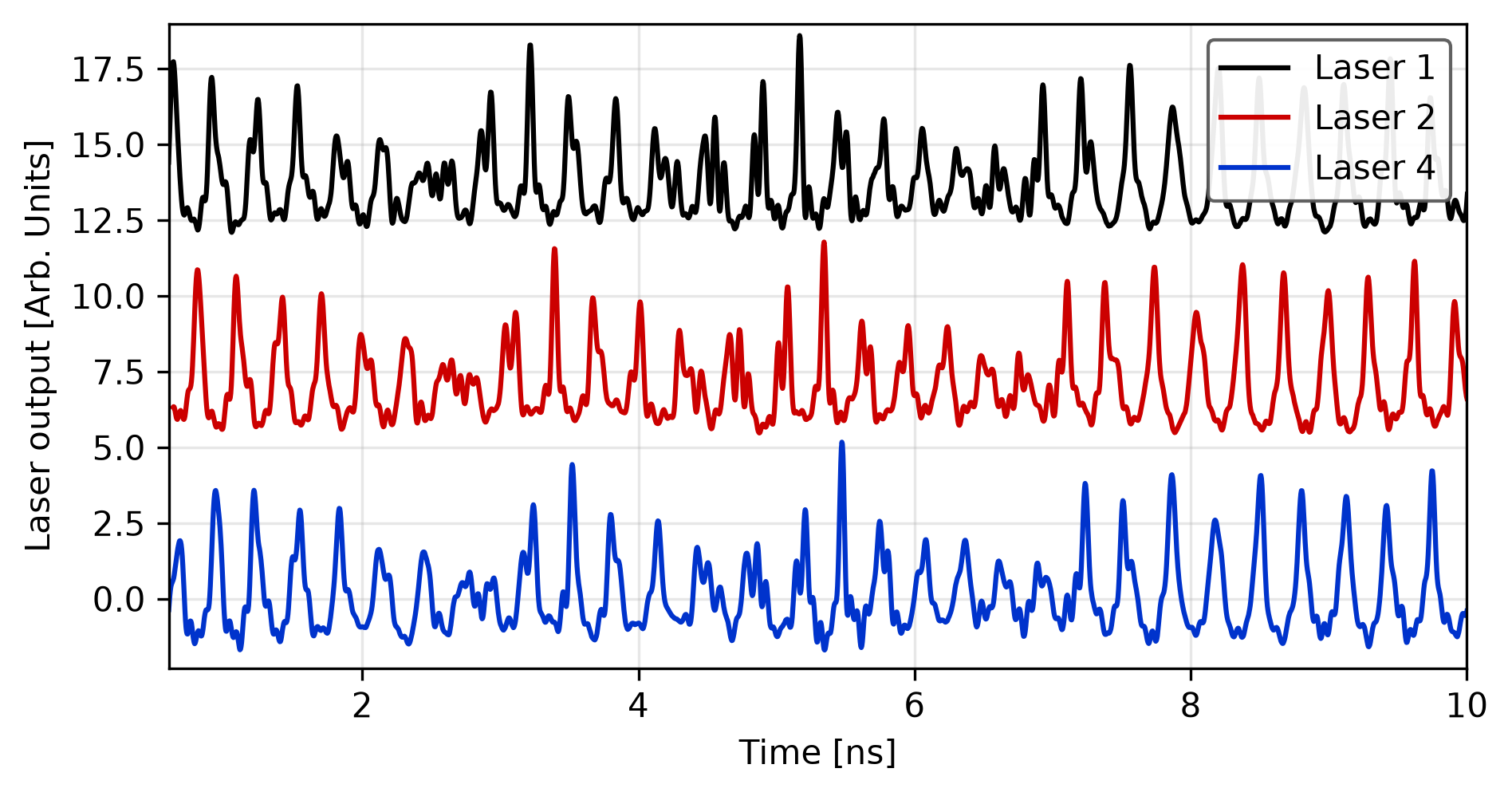}
\vspace{-0.8cm}
\caption{Short segment of the experimental time series recorded from three synchronized semiconductor lasers. In this case, the lasers exhibit synchronized chaotic dynamics with a small time lag.}
\label{fig:Experimental_TS}
\end{figure}

\begin{figure}[ht]
\centering
\includegraphics[width=1.015\linewidth]{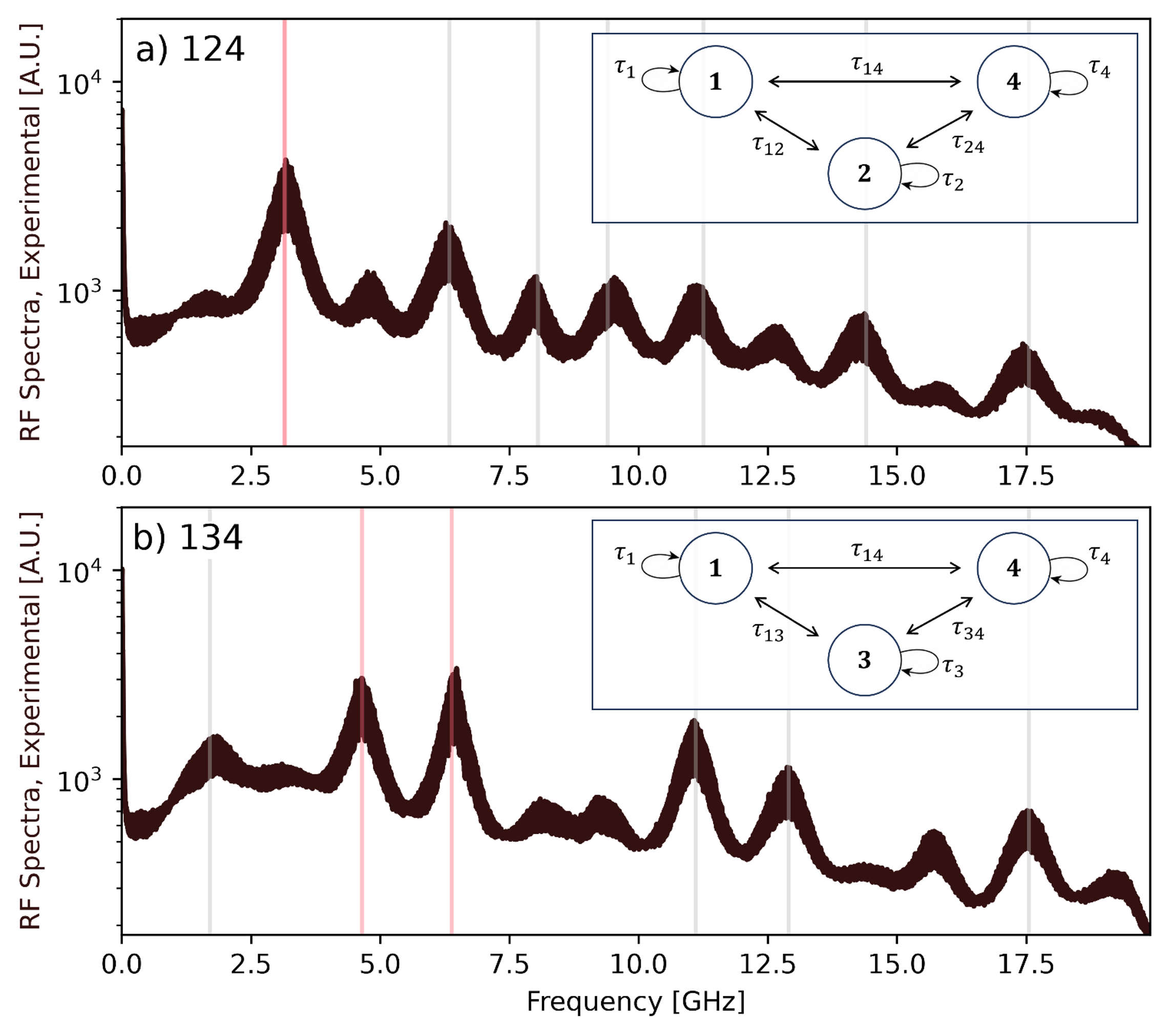}
\vspace{-0.7cm}
\caption{Experimental RFS of laser 1 for the (a) 124 and (b) 134 three-laser configurations. Remarkably, replacing laser 2 with laser 3 leads to a substantial change in the spectral structure, despite the two configurations differing by only a single laser. The insets show the setup of each configuration, where $\tau_i$ is the delay time, and $\tau_{ij}=(\tau_i+\tau_j)/2$.}
\label{fig:Experimental_FFTs}
\vspace{-0.1cm}
\end{figure}

A representative segment of the temporal series from three synchronized lasers is shown in Fig.~\ref{fig:Experimental_TS}. One of the most intriguing features of these chaotic dynamics is revealed in their radio-frequency spectra (RFS), which consistently exhibit a number of characteristic peaks whose physical origin needs to be thoroughly explored. In this work, we investigate this phenomenon in a network of three mutually coupled semiconductor lasers, with slight parameter inhomogeneities and non-identical delay times. 
To gain further insight into the origin of these spectral peaks, we study in detail two of the four three-laser configurations. Remarkably, the RFS obtained for each configuration differ substantially from one another, as shown in Fig. \ref{fig:Experimental_FFTs}, despite the fact that the overall network topology remains nearly unchanged. The different peaks observed in Fig.~\ref{fig:Experimental_FFTs}a) and Fig.~\ref{fig:Experimental_FFTs}b) suggest that the delay times, and in particular the mismatch between them, play a fundamental role in shaping the spectral characteristics of the synchronized chaotic dynamics.

\vspace{-0.2cm}

For each pair of lasers, the delay mismatch is defined as the absolute difference between their optical delays, according to
\begin{equation}
\delta_{ij}=|\tau_i-\tau_j|    
\end{equation}

Each delay mismatch defines a characteristic frequency according to $f_{ij}=1/\delta_{ij}$, while the associated harmonics correspond to integer multiples of $f_{ij}$. 
Table~\ref{table:tau_mismatch} lists the frequencies for all laser pairs together with their first two harmonics.

\begin{table}[ht]
\fontsize{8.9}{8.8}\selectfont
\centering
\caption{Delay mismatches $\delta_{ij}$ between each possible pair of lasers, with their corresponding frequencies $f_{ij}$ and harmonics. Optical delays are, for lasers 1 to 4 respectively: 82.25~ns, 82.60~ns, 82.71~ns, 82.88~ns.}
\vspace{0.1cm}
\begin{tabular}{ccccc}
\toprule
Laser pair $ij$ & $\delta_{ij}$ (ns) & $f_{ij}$ (GHz) & $2f_{ij}$ (GHz) & $3f_{ij}$ (GHz) \\
\midrule
12 & 0.35 & 2.86 & 5.71  & 8.57  \\
13 & 0.46 & 2.17 & 4.35  & 6.52  \\
14 & 0.63 & 1.59 & 3.17  & 4.76  \\
23 & 0.11 & 9.09 & 18.18 & 27.27 \\
24 & 0.28 & 3.57 & 7.14  & 10.71 \\
34 & 0.17 & 5.88 & 11.76 & 17.65 \\
\bottomrule
\end{tabular}
\label{table:tau_mismatch}
\end{table}


To investigate the hypothesis that the delay mismatch is responsible for the observed differences in the RFS, we employ a numerical model of the coupled laser network based on the Lang--Kobayashi equations \cite{LangKobayashi1980}. This model describes the dynamics of each semiconductor laser through a system of coupled differential equations governing the evolution of the complex electric field, $E(t)$, and the carrier population, $N(t)$. To account for the experimental configuration, the model incorporates a delayed self-feedback term representing optical feedback, as well as delayed coupling terms describing the mutual optical injection between the lasers. The equations employed here represent three identical single-mode lasers, subject to weak optical feedback (Lang-Kobayashi approximation) and weak optical mutual coupling. The equations are normalized such that the lasing threshold is at $\mu=1$.
\begin{equation}
\begin{aligned}
\frac{dE_i(t)}{dt}={}&
\frac{1}{2\tau_p}(1+i\alpha)(G_i-1)E_i(t)
+\sqrt{2\beta_{sp}}\xi_i(t) + \\
&+\eta E_i(t-\tau_i)e^{-i\omega_0\tau_i}
+\kappa \sum_{j\neq i}^{3}
\,E_j(t-\tau_{ij}) e^{-i\omega_0\tau_{ij}}
\end{aligned}
\label{eq:LK_E}
\end{equation}
\begin{equation}
   \dfrac{dN_i(t)}{dt}=\frac{1}{\tau_n}\left(\mu-N_i(t)-G_i|E_i|^2\right),
\label{eq:LK_N}
\end{equation}
\\
where ($i,j$) index the three lasers in each selected configuration. The nonlinear optical gain is given by $G_i=N_i/(1+\epsilon|E_i(t)|^2)$, where $\epsilon$ is the gain saturation coefficient. The parameter $\eta$ denotes the optical feedback strength, whereas $\kappa$ represents the optical injection strength. Owing to the symmetric experimental configuration, the injection and feedback strengths are assumed to be identical for all three lasers. The spontaneous emission factor is represented by $\beta_{sp}$, and $\alpha$ is the linewidth enhancement (Henry) factor, which couples the amplitude and phase of the optical field. The optical frequency, common to all lasers, is denoted by $\omega_0$, while $\mu$ is the normalized pump current. The parameter $\tau_i$ represents the external-cavity round-trip delay associated with optical feedback, and $\tau_{ij}$ is the propagation delay from laser $i$ to laser $j$, defined according to $\tau_{ij}=(\tau_i+\tau_j)/2$. The term $\xi(t)$ represents Gaussian white noise, uncorrelated in time, $\tau_p$ is the photon lifetime, and $\tau_n$ is the carrier lifetime. No frequency detuning between the lasers is considered. The terms in the exponentials in equation \ref{eq:LK_E} that correspond to the phases of feedback and coupling are set to zero. We have checked that similar results are obtained for other phases as long as the relative phases remain consistent with the experimental geometry. In addition, we focus exclusively on the synchronized chaotic regime, as it is the regime in which the delay-induced spectral features investigated here become most pronounced.
 

The equations are integrated numerically using the Heun method. The simulation parameters are summarized in Table~\ref{table:simulation_params}. Experimentally, the individual lasers exhibit slight parameter inhomogeneities, such as differences in the $\alpha$-factor and in the carrier lifetimes. To isolate the influence of the delay mismatches, the simulations assume identical parameter values for all lasers, except for the delay times. The parameter values are representative of the semiconductor lasers considered here.

\begin{table}[t]
\fontsize{8}{7.6}\selectfont
\centering
\caption{Parameters used in the simulations for all lasers.}
\label{table:simulation_params}
\begin{tabular}{clcc}
\toprule
Parameter& Brief description                          & Value used                   & Units \\
\midrule
$dt$             &  Integration step                    & 0.001                        & ns \\
$\mu$            &  Normalized pump current             & 1.2                          & -- \\
$\alpha$         &  Linewidth enhancement factor        & 2.65                         & -- \\
$\beta_{sp}$     &  Spontaneous noise coeff.            & $1\times10^{-5}$             & ns$^{-2}$ \\
$\tau_p$         &  Photon lifetime                     & 0.002                        & ns \\
$\tau_n$         &  Carrier lifetime                    & 1.45                         & ns \\
$\eta$           &  Feedback strength                   & 15                           & ns$^{-1}$ \\
$\kappa$         &  Injection strength                  & 15                           & ns$^{-1}$ \\
$\tau_i$         &  Feedback delay                      & Variable                     & ns  \\
$\tau_{ij}$      &  Coupling delays                     & $(\tau_i+\tau_j)/2$          & ns  \\
$\lambda_0$      &  Optical wavelength                  & 1554                         & nm \\
$\varepsilon$    &  Saturation parameter                & 0.05                         & --  \\
\bottomrule
\end{tabular}
\vspace{-0.2cm}
\end{table}

\begin{figure}[ht!]
\centering\includegraphics[width=1.002\linewidth]{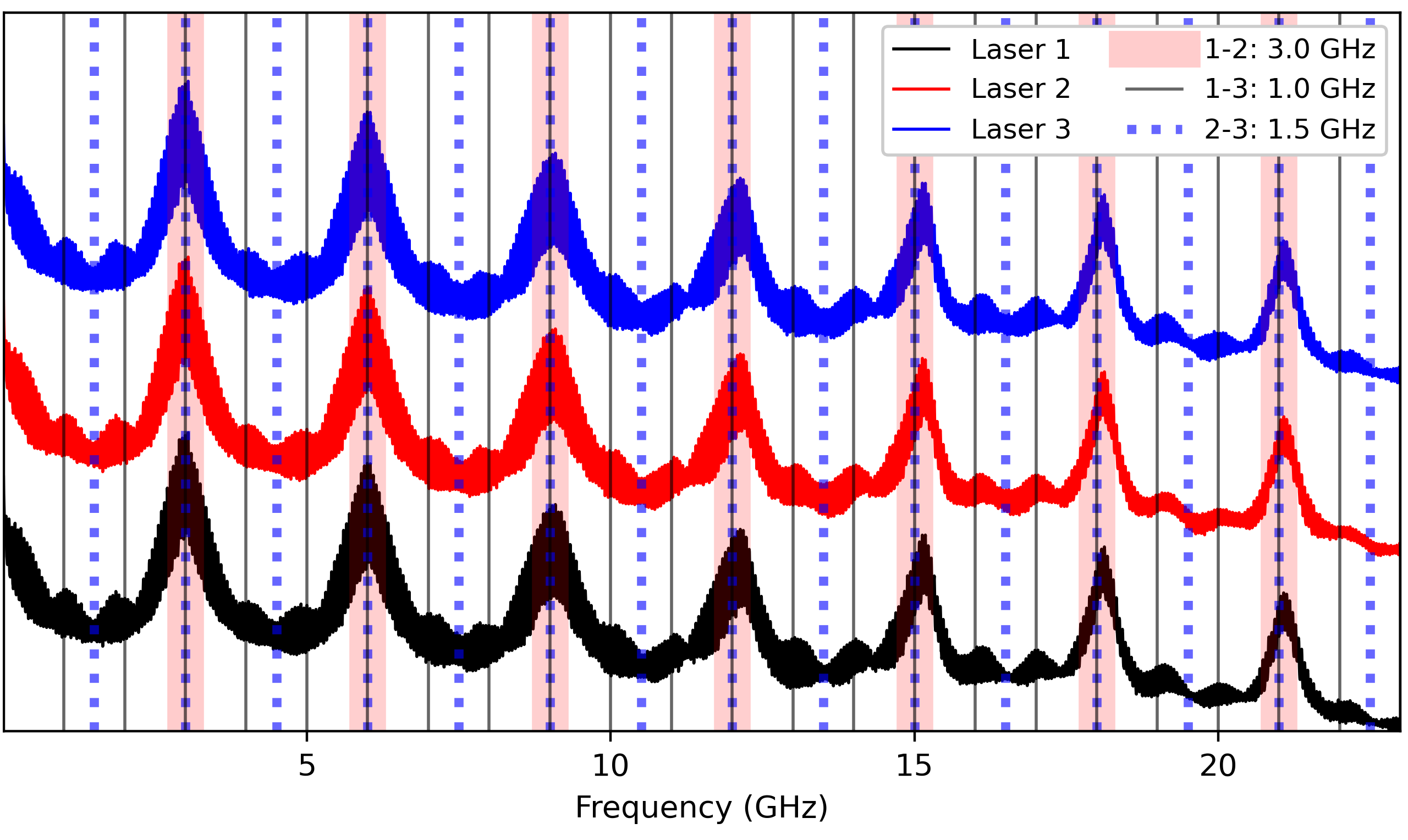}
\vspace{-0.7cm}
\caption{ RFS of simulated laser time series, vertically offset for clarity. Vertical lines indicate the theoretical frequencies associated with the delay mismatch between each pair of lasers, together with their higher-order harmonics. The characteristic spectral peaks are observed at frequencies where several of these theoretical lines coincide, indicating that the overlap of delay-mismatch harmonics determines the dominant features of the spectra. See more details for different delays at Visualization 1.}
\vspace{-0.2cm}
\label{fig:FFTs}
\end{figure}

As a reference case, we first simulate a system of three mutually coupled semiconductor lasers with optical feedback in the long-delay regime, using $\tau_i$ = 20, 20.333, and 21~ns for $i=1,2,3$, respectively. These delay times were chosen to yield characteristic frequencies with simple ratios: approximately 3 GHz (laser pair 1--2, $\delta_{12}=0.333$~ns), 1 GHz (pair 1--3, $\delta_{13}=1$~ns), and 1.5 GHz (pair 2--3, $\delta_{23}=0.667$~ns). In Fig.~\ref{fig:FFTs}, the positions of these frequencies and their harmonics are indicated by vertical lines with different colors and line styles for clarity (see the figure legend). Fig.~\ref{fig:FFTs} illustrates that the dominant spectral peaks occur at frequencies where several of these lines overlap, namely at integer multiples of 3~GHz, which corresponds to the least common multiple of the three characteristic frequencies. Furthermore, the RFS of the three lasers are identical, indicating that the synchronized lasers exhibit essentially the same frequency content. Therefore, for all the other figures presented in this work, only the spectrum of laser 1 is considered, which simplifies the comparison between the numerical simulations and the experimental results without loss of generality.


\begin{figure}[ht]
\centering
\includegraphics[width=1.016\linewidth]{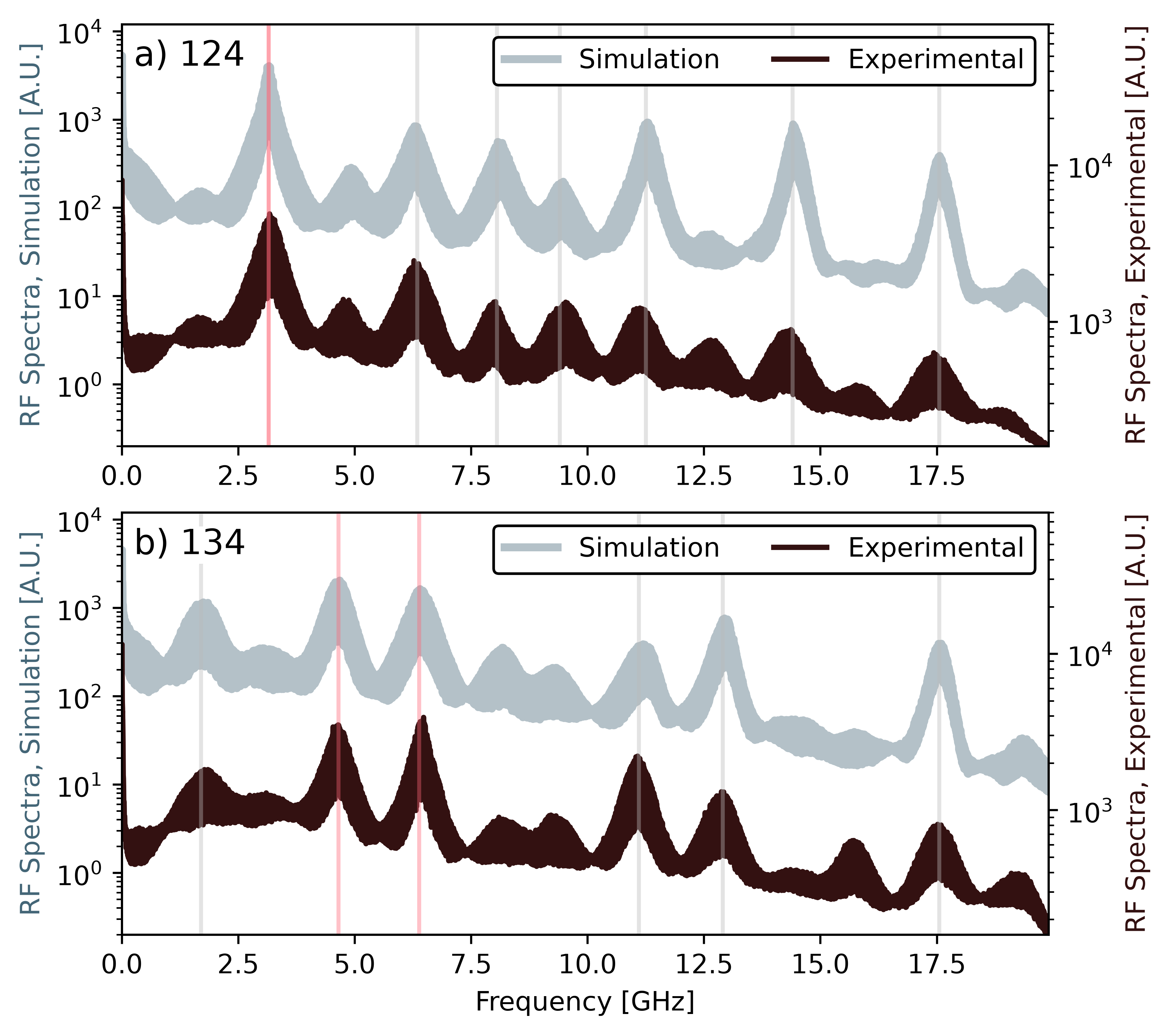}
\vspace{-0.7cm}
\caption{RFS of the simulated (light gray) and experimental (black) synchronized time series for the (a) 124 and (b) 134 three-laser configurations (only for laser 1). Consistent with the experimental observations, replacing laser 2 with laser 3 leads to significant modifications of the spectral structure, despite the two configurations differing only in the identity of a single laser.}
\vspace{-0.2cm}
\label{fig:Simulated_FFTs}
\end{figure}

Following the same approach, and encouraged by the clear spectral distribution for the example case in Fig.~\ref{fig:FFTs}, we now incorporate the experimentally measured delay times $\tau_i$ into the numerical model. This allows us to evaluate the influence of the actual delay mismatches present in the experimental setup on the resulting RFS. We analyze the same configurations presented in Fig.~\ref{fig:Experimental_FFTs}, using the experimentally measured delays while keeping the remaining parameters unchanged. The numerical results in Fig.~\ref{fig:Simulated_FFTs} reveal that the positions of the peaks in the RFS are highly sensitive to the values of the delay times. Even relatively small variations in $\tau_i$ lead to noticeable shifts in the spectral peaks, highlighting the critical role played by the delay mismatch in shaping the RFS. Most importantly, the simulated spectra reproduce the experimental peak structure to a large extent, with all the dominant spectral features appearing at the same frequencies as in the measurements. While the positions of the frequency peaks are well reproduced in the numerical simulations, the amplitude of the peaks beyond $\sim$7~GHz differs in the numerical results. This difference in the amplitude of the higher frequency peaks could be due to the omission of parameter mismatches between the lasers, but lies beyond the scope of this manuscript. These results provide strong evidence that the characteristic structure of the RFS is primarily governed by the relative differences between the time delays rather than by other experimental features or parameter variations. 

In practice, the spectral peak positions for configurations 124 and 134 have some trends in common. This is because the largest delay mismatch $\delta_{ij}$ corresponds to lasers 1 and 4,  which gives rise to the lowest characteristic frequency. Since both configurations include lasers 1 and 4, the resulting RFS exhibit a common comb of peaks with a spacing of $f_{14}=1/\delta_{14}$. For instance, the main peak in Fig.~\ref{fig:Experimental_FFTs}a) at around 3~GHz is a combination of $f_{12}=2.86$~GHz, $2f_{14}=3.17$~GHz and $f_{24}=3.57$~GHz, whereas in Fig.~\ref{fig:Experimental_FFTs}b) the first peak at around 4.5~GHz results from the combination of frequencies $2f_{13}=4.35$~GHz and $3f_{14}=4.76$~GHz, and the second peak at around 6.3~GHz results from the combination of frequencies $f_{34}=5.88$~GHz and $3f_{13}=6.52$~GHz (see Table~\ref{table:tau_mismatch} for more details). The difference between the two configurations arises from the intermediate laser, whose delay introduces additional frequency components that selectively enhance specific peaks of this underlying comb, effectively acting as a modulation envelope. More specifically, the spectral peaks near multiples of $f_{14}$ shift toward the characteristic frequencies associated with the remaining delay mismatches, namely $f_{12}$, $f_{13}$, $f_{24}$ and $f_{34}$. As a consequence, while the fundamental spectral structure is determined by the largest delay difference, the relative amplitudes of the peaks are shaped by the additional delay combinations involving the intermediate laser. At frequencies where no exact overlap exists, the spectral peaks tend to align with the closest characteristic frequencies or their harmonics, reflecting the structure imposed by the delay mismatches.

\begin{figure}[ht]
\centering
\includegraphics[width=1.011\linewidth]{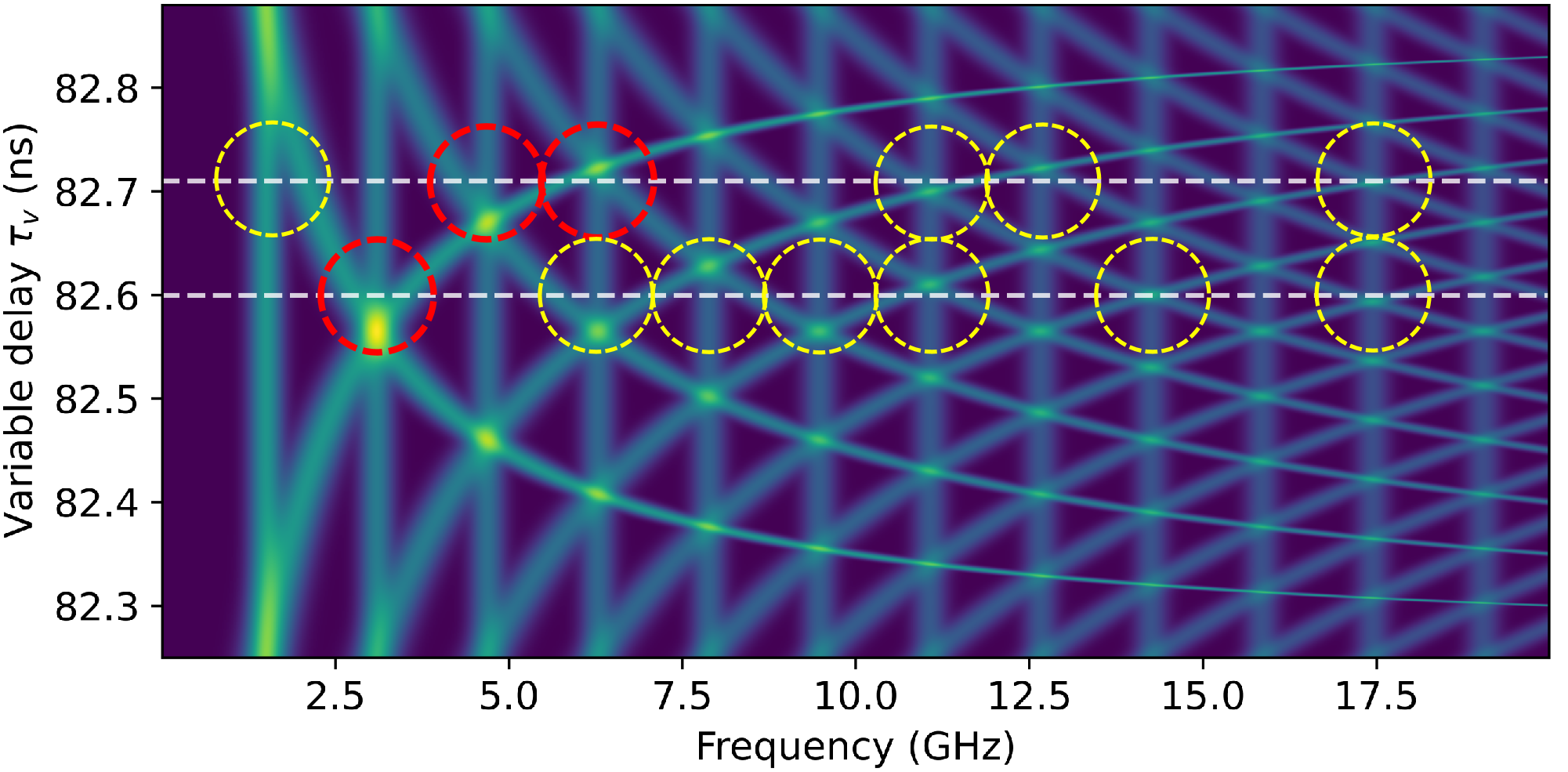}
\vspace{-0.6cm}
\caption{Predicted harmonic-crossing, where color code indicates the strength of the RF spectra, with fixed delays at 82.25~ns and 82.88~ns, and a third variable delay, $\tau_v$. Hyperbolic lines: prediction of RFS peaks between the variable laser and lasers 1 and 4, $f_{1v}^m=m/|\tau_1-\tau_v|$ and $f_{4v}^m=m/|\tau_4-\tau_v|$, where $m$ is an integer. Vertical lines: prediction of RFS peaks between lasers 1 and 4, $f_{14}^m=m/|\tau_1-\tau_4|$. Red circles indicate the positions of the main experimental peaks, whereas yellow circles indicate the secondary peaks.}
\vspace{-0.4cm}
\label{fig:spectra_heatmap}
\end{figure}

As a final verification, Fig.~\ref{fig:spectra_heatmap} presents a heatmap of the frequencies associated with all possible delay mismatches and their corresponding harmonics. The two extreme delays are fixed at $\tau_1$ and $\tau_4$, while the horizontal white lines indicate the positions of the intermediate laser. The hyperbolic curves represent the frequencies associated with the delay mismatches between the variable laser and lasers 1 and 4, namely $\delta_{1n}$ and $\delta_{4n}$, whereas the vertical lines correspond to the mismatch $\delta_{14}$. For the two laser configurations investigated, 124 and 134, the experimentally observed spectral peaks coincide with the regions where these frequency families exhibit the strongest overlap. The dominant peaks identified in Fig.~\ref{fig:Simulated_FFTs} are highlighted by red circles.
 
A systematic scan covering all possible delay values for three lasers is provided in Visualization 1. In particular, peaks separated by 1.59~GHz, corresponding to the inverse of the delay difference between $82.25\,\mathrm{ns}$ and $82.88\,\mathrm{ns}$, continuously shift and redistribute in order to match the spectral components associated with the varying delay $\tau_v$. This behaviour illustrates how the complete RFS results from the superposition and mutual accommodation of the frequency combs generated by the different delay mismatches present in the coupled laser network.
\vspace{-0.3cm}

In conclusion, we have investigated the influence of delay-time mismatches on the RFS of mutually coupled semiconductor lasers and demonstrated that these mismatches determine the characteristic frequencies governing the positions of the dominant spectral peaks. The proposed interpretation is supported by both experimental measurements and numerical simulations, which exhibit a clear agreement in terms of the location and relative prominence of the observed peaks. The same principle of spectral alignment induced by delay mismatches is expected to extend to larger laser networks, where additional delay combinations may make the RF spectrum even more selective than in the three-laser case \cite{SPIE_proceedings}. Such delay-mismatch-induced spectral selectivity could be detrimental for applications of laser networks in chaos-based communications, where broadband and less structured chaotic spectra are often desirable. Conversely, it may be beneficial for information-processing applications, where controllable oscillatory components and reproducible spectral structure can provide useful dynamical resources \cite{Goldmann2023}.


\vspace{.1cm}

\begin{backmatter}
\bmsection{Funding} S.T.S. acknowledges the support of Agencia de Gestió d’Ajuts Universitaris i de Recerca (AGAUR, Joan Oró FI 2025). INFOLANET project (No. PID2022-139409NB-I00), funded by MCIN/AEI/10.13039/501100011033. C.M. acknowledges project PID2024-160573NB-I00 funded by MICIU/AEI/10.13039/501100011033 and FEDER, UE and project FA8655-24-1-7022 funded by EOARD.
\vspace{-0.1cm}
\bmsection{Acknowledgments} We want to thank Adam C. Mackowiak-Pellon for stimulating discussions.  
\vspace{-0.1cm}
\bmsection{Disclosures} The authors declare no conflicts of interest.
\vspace{-0.1cm}
\bmsection{Data availability} Data underlying the results presented in this paper are available from the authors upon reasonable request.

\end{backmatter}

\bibliography{references}

\bibliographyfullrefs{references}

\end{document}